\documentclass[prl, superscriptaddress, showpacs, floatfix, nobalancelastpage, twocolumn]{revtex4}

\usepackage[]{graphicx}
\usepackage{amsmath}
\usepackage{amsfonts}

\def\ket#1{| #1 \rangle}

\def\cS{{\cal S}}
\def\cE{{\cal E}}
\def\Tr{{\mathrm{Tr}}}
\newcommand{\eend}      {\hspace{\stretch{1}}\rule{1ex}{1ex}}

\pacs{03.65.Yz, 03.65.Ta, 03.67.-a}
\date{\today}

\begin{document}

\title{Objective properties from subjective quantum
  states: Environment as a witness}

\author{Harold Ollivier}
\affiliation{Theoretical Division, LANL, MS-B213, Los Alamos, NM  87545, USA}
\affiliation{INRIA - Projet Codes, BP 105, F-78153 Le Chesnay, France}
\author{David Poulin}
\affiliation{Theoretical Division, LANL, MS-B213, Los Alamos, NM  87545, USA}
\affiliation{Institute for Quantum Computing, University of Waterloo, ON Canada, N2L 3G1}
\author{Wojciech H. Zurek}
\affiliation{Theoretical Division, LANL, MS-B213, Los Alamos, NM  87545, USA}

\begin{abstract}
  We study the emergence of objective properties in open quantum
  systems. In our analysis, the environment is promoted from a passive
  role of reservoir selectively destroying quantum coherence, to an
  active role of amplifier selectively proliferating information about
  the system. We show that {\em only preferred pointer states of the
  system can leave a redundant and therefore easily detectable imprint on
  the environment}. Observers who---as it is almost always the
  case---discover the state of the system indirectly (by probing a
  fraction of its environment) will find out only about the
  corresponding pointer observable. Many observers can act in this
  fashion independently and without perturbing the system: they will
  agree about the state of the system. In this operational sense, {\em
  preferred pointer states exist objectively.}
\end{abstract}

\maketitle

The key feature distinguishing the classical realm from the quantum
substrate is its objective existence. Classical states can be found
out through measurements by an initially ignorant observer without
getting disrupted in the process. By contrast, an attempt to discover
the state of a quantum system through a direct measurement generally
leads to a collapse~\cite{Boh28a, Neu32a, Dir47a}: after a
measurement, the state will be what the observer finds out it is, but
not---in general---what it was before. Thus, it is difficult to claim
that quantum states exist objectively in the same sense as their
classical counterparts~\cite{Bohr27,EPR35a,FP00a}.

It is by now widely appreciated that decoherence, caused by persistent
monitoring of a system by the environment, can single out a preferred
set of states. In simplest models, such pointer states~\cite{Zur93b,
Zur98a, PZ01a, Zur03a, GJKK96a} are (often degenerate) eigenstates of
the pointer observable which commutes with the system-environment
interaction~\cite{Zur82a}. This concept can be generalized using the
predictability sieve: only pointer states evolve predictably in spite
of the openness of the system~\cite{Zur93b, PZ01a, Zur03a}.  They
exist in the sense that in absence of any perturbations---save for the
monitoring by the environment---they or their dynamically evolved
descendants will continue to faithfully describe the system. Thus,
when an observer knows what are the pointer states, he can learn which
of them represents the system without perturbing it. However, when an
observer ignorant of what pointer states are attempts to find out the
state of the system directly, he still faces, even in the presence of
decoherence, the danger of collapsing its wave packet.

Here, we build on the idea that a {\em direct} measurement of the
system is not how observers gather data about the Universe: rather, a
vast majority (if not all) of our information is obtained {\em
indirectly} by probing a small fraction of the
environment~\cite{Zur93b, Zur98a, Zur03a}. One may think that
this twist in the story can be accounted for by adding a few links to
the von Neumann chain~\cite{Neu32a}, but this is not the case: we
shall show that the monitoring environment acquires information about
the system {\em selectively}. More importantly, this selective
spreading of information through the environment---in essence
``quantum Darwinism''~\cite{Zur03a}---accounts for the objective
existence of some preferred quantum states: by probing the system
indirectly, hence without perturbing it, many independent observers
can obtain reliable information, but only about the pointer states.

This letter is organized as follows: first, we introduce our
operational definition of objectivity. We then state necessary and
sufficient conditions for the objective existence of an observable in
the context of {\em einselection} (environment-induced
superselection). Next, these requirements are translated into an
information theoretic framework, and proven to imply a {\em unique}
observable: the usual pointer observable. This is our key
result. Finally, we show that, because of quantum Darwinism,
information about pointer states is robust and, hence, objective.

{\em An operational definition of objectivity} for a property of a
quantum system should not rely on pre-existence of an underlying
reality as it is presumed in the classical setting. Rather, we demand
that an objective property of the system of interest should be\\ ({\em
i}\,)~simultaneously accessible to many observers,\\ ({\em ii}\,)~who
should be able to find out what it is without prior knowledge and\\
({\em iii}\,)~who should arrive at a consensus about it without prior
agreement.\\ As we already mentioned, the collapse of the wave packet
following a direct measurement generally precludes this. However, when
the system of interest $\cS$ interacts with an environment $\cE$
composed of many subsystems, $\cE = \bigotimes_{k=1}^N \cE_k$, the
situation changes dramatically. When a property leaves a {\em
complete} and {\em redundant} imprint on the environment, all three
criteria are satisfied: many copies are available, hence simultaneous
accessibility ({\em i}\,) is straightforward.  Moreover fractions of
the environment can be measured without perturbing either $\cS$ or the
rest of $\cE$.  Therefore, ignorant observers can vary their
measurements independently, corroborate their own results and arrive at
a common description of properties of the system. Hence, owing to
redundancy, prior knowledge ({\em ii}) is not necessary to ({\em iii})
reach consensus. The existence of an objective property requires the
presence of its {\em complete} and {\em redundant} imprint in the
environment as necessary and sufficient conditions. Our approach will
focus on the study of the correlations between parts of the
environment and the system of interest.

\paragraph{Information theoretical framework.}
A natural way to characterize such correlations is to use the mutual
information $I(\sigma : \mathfrak e)$ between an observable $\sigma$
of $\cS$ and a measurement $\mathfrak e$ on $\cE$. In short, $I(\sigma
: \mathfrak e)$ measures one's ability to predict the outcome of
measurement $\sigma$ on $\cS$ after having ``looked at the
environment'' through $\mathfrak{e}$. For a given density
matrix $\rho^{\cS\cE}$ of $\cS\otimes \cE$, the measurement results
are random variables characterized by a joint probability distribution
\begin{equation}
p( \sigma_i, \mathfrak{e}_j) = \Tr \{( \sigma_i \otimes
\mathfrak{e}_j) \rho^{\mathcal S \mathcal E}\},
\end{equation}
where $ \sigma_i$ and
$\mathfrak{e}_j$ are the spectral projectors of observables $\sigma$
and $\mathfrak e$. By definition, the mutual information is the
difference between the initially missing information about $\sigma$
and the remaining uncertainty about $\sigma$ when $\mathfrak e$ is
known~\cite{CT91a}. Using Shannon entropy as a measure of missing
information, $H(\sigma) = -\sum_i p(\sigma_i) \log p(\sigma_i)$ and
$H(\sigma,\mathfrak e) = -\sum_{i,j} p(\sigma_i, \mathfrak e_j) \log
p(\sigma_i, \mathfrak e_j)$, the mutual information is 
\begin{equation}
I(\sigma : \mathfrak{e}) = H( \sigma) + H(\mathfrak{e}) -
H(\sigma, \mathfrak{e}).
\end{equation}

The information about observable $\sigma$ of $\cS$ that can be
optimally extracted from $m$ environmental subsystems is
\begin{equation}
\hat I_m(\sigma) =
\max_{\{\mathfrak{e}\in \mathfrak M_m\}}I(\sigma : \mathfrak{e})
\label{eq:I_m}
\end{equation}
where $\mathfrak M_m$ is the set of all measurements on those $m$
subsystems. In general, $\hat I_m(\sigma)$ will depend on which
particular $m$ subsystems are considered. For simplicity, we will
assume that any {\em typical} $m$ environmental subsystems yield
roughly the same information. This may appear to be a strong
assumption, but, as we discuss bellow, relaxing it does not affect our
main conclusions. By setting $m$ to the total number $N$ of subsystems
of $\cE$, we get the information content of the entire
environment. Then,
\begin{equation}
\hat I_N(\sigma) \approx H(\sigma)
\label{eq:transfer}
\end{equation} 
expresses the {\em completeness} prerequisite for objectivity: all (or
nearly all) missing information about $\sigma$ must be in principle
obtainable from all of $\cE$.

\begin{figure*}[tb] 

\center \includegraphics[width=5.8in]{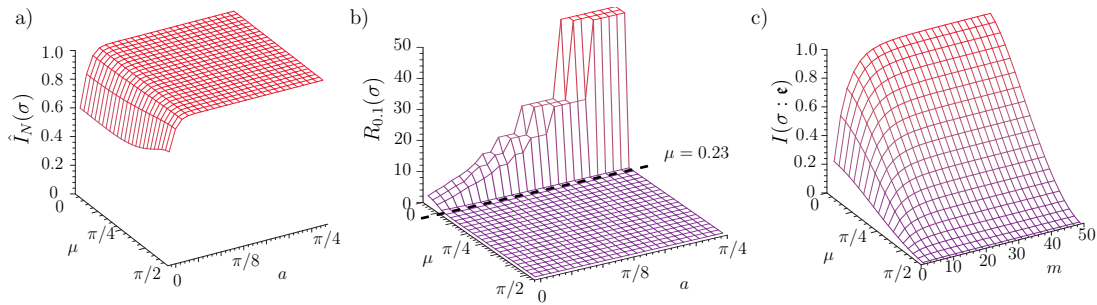}

\caption{Quantum Darwinism can be illustrated using a model introduced
  in~\cite{Zur82a}. The system $\cS$, a spin-$\frac 12$ particle,
  interacts with $N=50$ two-dimensional subsystems of the environment
  through $\hat H^{\cS\cE} = \sum_{k=1}^N g_k \sigma_z^{\cS}
  \otimes\sigma_y^{\cE_k}$ for a time $t$. The initial state of
  $\mathcal S\otimes\cE$ is $\frac{1}{\sqrt 2}(\ket 0 + \ket 1)
  \otimes \ket 0 ^{\cE_1}\otimes\ldots\otimes\ket 0^{\cE_N}$. All the
  plotted quantities are function of the system's observable
  $\sigma(\mu) = \cos(\mu)\sigma_z + \sin(\mu)\sigma_x$, where $\mu$
  is the angle between its eigenstates and the pointer states of
  $\cS$---here the eigenstates of $\sigma_z^\cS$. {\bf a)}~Information
  acquired by the optimal measurement $\mathfrak e$ on the whole
  environment, $\hat I_N(\sigma)$, as a function of the inferred
  observable $\sigma(\mu)$ and the action $a_k = g_k t = a$ for all
  $k$. A large amount of information is accessible in the {\em whole}
  environment for any observables $\sigma(\mu)$ except when the
  interaction action $a_k$ is very small. Thus, complete imprinting of
  an observable of $\cS$ in $\cE$ is not sufficient to claim
  objectivity. {\bf b)}~Redundancy of the information about the system
  as a function of the inferred observable $\sigma(\mu)$ and the
  action $a_k = g_k t = a$. It is measured by
  $R_{\delta=0.1}(\sigma)$, which counts the number of times 90\% of
  the total information can be ``read off'' independently by measuring
  distinct fragments of the environment. For all values of the action
  $a_k = g_k t =a$, redundant imprinting is sharply peaked around the
  pointer observable. Redundancy is a very selective criterion. The
  number of copies of relevant information is high only for the
  observables $\sigma(\mu)$ falling inside the theoretical bound (see
  text) indicated by the dashed line.  {\bf c)}~Information about
  $\sigma(\mu)$ extracted by an observer restricted to local random
  measurements on $m$ environmental subsystems (e.g.\ $\mathfrak e =
  \mathfrak e^{\cE_1}\otimes\ldots \otimes\mathfrak e^{\cE_m}$ where
  each $\mathfrak e^{\cE_k}$ is chosen at random).  The interaction
  action $a_k = g_k t $ is randomly chosen in $[0,\pi/4]$ for each
  $k$. Because of redundancy, pointer states---and only pointer
  states---can be found out through this far-from-optimal measurement
  strategy. Information about any other observable $\sigma(\mu)$ is
  restricted by our theorem to be equal to the information brought
  about it by the pointer observable $\sigma_z^\cS$,
  Eq.~(\ref{bound_th}).}
\label{figure}
\end{figure*}

However, as a consequence of basis ambiguity~\cite{Eve57a, Zur82a},
information about many observables $\sigma$ can be deduced by an
appropriate measurement on the entire environment. Therefore,
completeness, Eq.~(\ref{eq:transfer}), while a prerequisite for
objectivity, is not a very selective criterion (see Fig.~\ref{figure}a
for illustration).  To claim objectivity, it is not sufficient to have
a complete imprint of the candidate property of $\cS$ in the
environment. There must be many copies of this imprint that can be
accessed independently by many observers: {\em information must be
redundant}.

\paragraph{Redundancy and its consequences.}
To obtain a measure of redundancy, one must count the number of copies
of the information about $\sigma$ present in $\cE$. Redundancy is thus
quantified by the number of disjoint subsets of $\mathcal E$
containing almost all---all but a fraction $\delta$---of the
information about $\sigma$ available from the entire $\cE$:
\begin{equation}
R_\delta(\sigma) = {N}/{m_\delta(\sigma)}.
\label{def_red} 
\end{equation}
Above $m_\delta(\sigma)$ is the smallest number $m$ of typical
environmental subsystems that contain almost all the information about
$\sigma$, i.e.\ $\hat I_m(\sigma) \geq (1-\delta)\hat
I_N(\sigma)$).

The key question now is: What is the structure of the set $\mathcal O$
of observables that are {\bf completely}, $I_N(\sigma) \approx
H(\sigma)$, and {\bf redundantly}, $R_\delta(\sigma) \gg 1$ with
$\delta \ll 1$, imprinted on the environment? The answer is provided
by the following theorem.
  
\noindent {\bf Theorem:}~{\it The set $\mathcal O$ is characterized by
a unique observable $\pi$, called by definition the {\bf maximally
refined observable}: the information $\hat I_m(\sigma)$ about any
observable $\sigma$ in $\mathcal O$ obtainable from a fraction of
$\cE$ is equivalent to the information about $\sigma$ that can be
obtained through its correlations with the maximally refined
observable $\pi$:
\begin{equation}
\hat I_m(\sigma) = I(\sigma:\pi)
\label{bound_th}
\end{equation}
for $m_\delta(\pi) \leq m \ll N$.} 

\paragraph{Outline of the proof for perfect records, $\delta = 0$.}
Let $\sigma^{(1)}$ and $\sigma^{(2)}$ be two observables in $\mathcal
O$ for $\delta = 0$. Since $\sigma^{(1)}$ and $\sigma^{(2)}$ can be
inferred from two disjoint fragments of $\cE$, they must
commute. Similarly, let $\mathfrak{e}^{(1)}$
(resp. $\mathfrak{e}^{(2)}$) be a measurement acting on a fragment of
$\cE$ that reveals all the information about $\sigma^{(1)}$
(resp. $\sigma^{(2)}$) while causing minimum disturbance to
$\rho^{\cS\cE}$. Then, $\mathfrak{e}^{(1)}$ and $\mathfrak{e}^{(2)}$
commute, and can thus be measured {\em simultaneously}. This combined
measurement gives complete information about $\sigma^{(1)}$ {\em and}
$\sigma^{(2)}$. Hence, for any pair of observables in $\mathcal O$, it
is possible to find a more refined observable which is also in
$\mathcal O$. The maximally refined observable $\pi$ is then obtained
by pairing successively all the observables in $\mathcal O$. By
construction $\pi$ satisfies Eq.~(\ref{bound_th}) for any $\sigma$ in
$\mathcal O$.\eend

Note that the Theorem does not guarantee the existence of a {\em non
trivial} observable $\pi$: when the system does not properly correlate
with $\cE$, the set $\mathcal O$ will only contain the identity
operator.

In fact, this Theorem can be extended to nearly perfect
records for assumptions satisfied by usual models of
decoherence~\cite{OPZ04a}. The proof is based on the
recognition that only the already familiar pointer observable can have
a redundant and robust imprint on $\cE$. The Theorem can be understood
as a consequence of the ability of the pointer states to persist while
immersed in the environment. This resilience allows the information
about the pointer observables to proliferate, very much in the spirit
of the ``survival of the fittest".

Two important consequences of this theorem follow. ({\em i}\,)~An
observer who probes only a fraction of the environment is able to find
out the state of the system as if he measured $\pi$ on $\cS$. ({\em
ii}\,)~Information about any other observable $\sigma$ of $\cS$ will
be inevitably limited by the available correlations existing between
$\sigma$ and $\pi$. In essence, our theorem proves the uniqueness of
redundant information, and therefore the selectivity of its
proliferation.

{\em Quantum Darwinism}---the idea that the perceived {\it classical
reality} is a consequence of the selective proliferation of
information about the system~\cite{Zur03a}---is consistent with
previous approaches to einselection, such as the predictability sieve,
but goes beyond them. The existence of redundant information about the
system, induced by specific interactions with the environment,
completely defines what kind and how information can be retrieved from
$\cE$: Eq.~(\ref{bound_th}) shows that the most efficient strategy for
inferring $\sigma$ consists in estimating $\pi$ first, and deducing
from it information about $\sigma$. It also explains the emergence of
a consensus about the properties of a system. Observers that gain
information about $\pi$---the only kind of information available in
fragments of $\cE$---will agree about their conclusions: their
measurement results are directly correlated with $\pi$, and are
therefore correlated with each other.  Hence, observers probing
fractions of the environment can act {\em as if} the system had a
state of its own---an {\em objective} state (one of the eigenstates of
${\pi}$). By contrast, such consensus cannot arise for superpositions
of pointer states, e.g.\ Schr\"odinger cats, since information about
them can only be extracted by probing the whole environment, and thus
cannot be obtained independently by several observers.  Objectivity
comes at the price of singling out a preferred observable of $\cS$
whose eigenstates are redundantly recorded in $\cE$. Cloning of quanta
is not possible~\cite{WZ82a}, but amplification of a preferred
observable happens almost as inevitably as decoherence, and leads to
objective classical reality. The impossibility of cloning and the
capacity for amplification imply selection---Darwinian ``survival of
the fittest''.

\paragraph{Emergence of objectivity exemplified.}
In Figure~\ref{figure}b, we show for a specific model the redundancy,
$R_{\delta = 0.1}$, as a function of the inferred observable
$\sigma(\mu)$ (whose eigenstates are tilted by an angle $\mu$ from the
pointer ones) and of the interaction action, $a_k = g_k t$ ($\sin(a_k)$
characterizes the strength of the correlations between $\cS$ and
$\cE_k$). By carefully tracking all orders of $\delta$ in
Eq.~(\ref{bound_th}), one can show that the existence of a complete
and redundant imprint of observable $\sigma(\mu)$ in the environment
requires $H_2(\cos^2 \frac\mu 2) \leq \delta$, where $H_2(p) = -[p\log
p + (1-p)\log (1-p)]$. Inserting the actual values of the parameters
chosen for our simulation, the above equation indicates that only
observables with $|\mu|<0.23$ leave a redundant imprint on the
environment: the objective properties of the system are unique. This
bound is in excellent agreement with our numerical
results. Surprisingly, and as confirmed by our simulation, the
interaction action $a_k$ only plays a role in setting the value of the
redundancy at its maximum, but does not affect the selectivity of our
criterion. Which observable becomes objective is largely decided by
the structure of the interaction Hamiltonian, (i.e.\ the set of
pointer states), but not by its details such as strength and duration
of the interaction. This ensures the stability of the pointer
observable deduced from redundancy.

\paragraph{Robustness of information.}
Objective information must be extractable through
``realizable''---hence, not necessarily optimal---measurements for
many observers to arrive at an operational consensus about the state
of a system. For instance, human eyes can only measure photons {\em
separately}, yet we can still learn about the position of
objects. This issue is considered for our model in
Fig.~{\ref{figure}c}. Here, even local (i.e.\ spin by spin) {\em
random} measurements eventually acquire the entire information
available in $\cE$ about the pointer states. Though surprising, this
result naturally follows from quantum Darwinism and the fact that high
redundancy protects information against a wide range of errors. Almost
any observable of $\cS$ is completely imprinted on the environment
(see Fig.~\ref{figure}a). However, as our theorem establishes and
Fig.~\ref{figure}b illustrates, only the observables which are
``close'' to the maximally refined pointer observable $\pi =
\sigma_z^\cS$, can be imprinted redundantly in the
environment. Therefore only information about pointer states can
tolerate errors, i.e.\ can be extracted by non-optimal
measurements. In short, {\em not only is the information about the
pointer observable easy to extract from fragments of the environment,
it is impossible to ignore!}

Clearly, for the emergence of objective properties, it is much more
important to know that $R_\delta(\sigma) \gg 1$ than to know its
precise value: the whole idea of redundancy is that it allows one to
be sloppy in decoding the message and still ``get it right''. This is
our proposed explanation for the robustness of the classical. Thus,
the essence of our conclusions is unaffected by the assumption (see
Eq.~(\ref{eq:I_m})) of even spreading of information in the
environment: our result depends only on the existence of multiple
records of the same information in disjoint fragments of the
environment, not on what these fragments are.

Similarly, $R_\delta(\sigma)$ depends on the tensor product of $\cE$
into subsystems, which is {\em a priori} arbitrary. However, several
considerations suggest {\em locality} as guide line for the definition
of elementary subsystems. For instance, our access to the information
content of the environment is restricted by the fundamental
Hamiltonians of nature which are local. Moreover, various observers
occupy, and therefore monitor, different spatial regions. Hence, an
operational notion of redundancy should reflect spreading of
information in space. Again, details of the partition are not
important as we are only interested in the typical behavior of
redundancy. These issues will be discussed in more details in a
forthcoming paper~\cite{OPZ04a}.

{\em Quantum Darwinism} capitalizes on some ideas that arose in the
context of decoherence and einselection, but goes beyond them in an
essential fashion.  Existence of records in the environment has been
noted before~\cite{Zur93b, GJKK96a, Zur98a, Hal99b}, and the
fact that it is easiest to find out about the pointer observable has
been also appreciated~\cite{DDZ01a}. Here, however, we have described
an even more dramatic turn of events---environment as a broadcast
medium--- which may seem fanciful until we realize that it describes
rather accurately what happens in the real world.  For instance, human
observers acquire all of their visual data by intercepting a small
fraction of their photon environment. An operational notion of
objectivity emerges from redundant information as it enables many
independent observers to find out the state of the system without
disturbing it. Furthermore, {\it objective} observables are
robust---insensitive to changes in the strategy through which the
environment is interrogated, as well as to variations of the strength
and duration of the interaction between $\cS$ and $\cE$, etc. One
might regard quantum Darwinism as a fully quantum implementation of
Bohr's idea~\cite{Bohr1958} about the role of amplification in the
transition from quantum to classical.

We thank R.~Blume-Kohout, D.~Dalvit and C.~Fuchs for stimulating discussions
and C.~Negrevergne for comments on the manuscript.  This research was
supported in part by NSA and ARDA.

\end{document}